\documentclass[preprint,showpacs,preprintnumbers,amsmath,amssymb]{revtex4}

\usepackage{graphicx}
\usepackage{dcolumn}
\usepackage{bm}

\begin{document}

\preprint{}

\title{Quantum and Fisher Information from the 
Husimi and Related Distributions}

\author{Paul B. Slater}%
\email{slater@kitp.ucsb.edu}
\affiliation{%
ISBER, University of California, Santa Barbara, CA 93106\\
}%
\date{\today}

\begin{abstract}
The two principal/immediate influences --- which we seek to interrelate
here --- upon the undertaking of this study are papers of 
\.Zyczkowski and S{\l}omczy\'nski (J. Phys. A  34, 6689 [2001]) 
and of Petz and Sud\mbox{\'a}r (J. Math. Phys. 37, 2262 [1996]).
In the former work, a metric (the Monge one, specifically) over
generalized Husimi distributions was employed to define a distance
between two arbitrary density matrices.  In the Petz-Sud\mbox{\'a}r 
work (completing a program of Chentsov), 
the quantum analogue of the (classically unique) 
Fisher information (montone) metric of a 
probability simplex was extended to define an uncountable infinitude of 
Riemannian (also monotone) metrics on the set of 
positive definite density matrices.
We pose here the 
questions of what is the specific/{\it unique} 
Fisher information metric for
the (classically-defined) Husimi distributions and how does it relate
to the {\it infinitude} of (quantum) metrics over the density matrices 
of Petz and Sud\mbox{\'a}r?
We find 
a highly proximate (small relative entropy) relationship
between the probability distribution (the quantum Jeffreys' prior) 
that yields quantum 
universal data compression, and that which (following Clarke and Barron) 
gives its classical counterpart.
We also investigate the Fisher information metrics 
corresponding to the {\it escort}
Husimi, positive-P and certain Gaussian
probability distributions, as well as, 
in some sense, the discrete Wigner {\it pseudoprobability}. 
The {\it comparative noninformativity} 
of prior probability distributions --- recently 
studied by Srednicki (Phys. Rev. A 71, 052107 
[2005]) --- formed by normalizing the volume elements 
of the various 
information metrics, is also discussed in our context.

\end{abstract}

\maketitle
\section{Introduction}
The two-level quantum systems (TLQS) are describable (nonclassically) 
 in terms of
$2 \times 2$ density matrices 
($\rho$) --- Hermitian nonnegative definite matrices of 
trace unity. These matrices can be parametrized by points in the unit ball
(Bloch ball/sphere \cite[p. 10244]{mosseri}) 
in Euclidean 3-space. 
On the other hand, the TLQS can be described in a {\it classical}
manner using a generalization of the Husimi distribution 
\cite{khusimi}
 \cite[sec. 4.1]{monge} (cf. \cite{guiasu,levine,bach,bach2,bach3,gibbons}).
``The Husimi function is a function on phase space, and takes only
non-negative values while the Wigner function can be negative and is usually
violently oscillating. Hence the Husimi function can be regarded as a
probability distribution in phase space, and its order of delocalization
can be a measure of chaoticity of quantum states'' \cite{sugita2}. 
(Note that the original Husimi distribution was defined only for density operators
in {\it separable} Hilbert space --- one which admits a countable orthonormal 
basis --- while the distribution to be studied here is
defined over a finite-dimensional Hilbert space.)

There is an (uncountable) {\it infinitude} 
\cite[sec. 16.7]{penrose} of
(quantum monotone) Riemannian 
metrics that can be attached to the Bloch ball of TLQS. 
Contrastingly, 
in the
classical
context of the Husimi distribution,
there is not an infinitude, but rather a {\it single} distinguished
(up to a constant multiple)
monotone Riemannian
metric --- the {\it Fisher information} metric
\cite{chentsov,campbell,frieden}. 
(``In the classical case, decision theory provides a unique montone
metric, namely, the Fisher information. In the quantum case, there are 
infinitely many monotone metrics on the state space'' 
\cite[p. 2672]{petzsudar}.)
So, it appears to be an question of obvious interest --- which we seek to 
address here --- of how one reconciles/deals
with this phenomenon of classical uniqueness and quantum non-uniqueness, 
as applied to essentially the {\it same} objects (that is, the TLQS).
\section{Monotone metrics}
The monotone metrics are all {\it stochastically monotone} \cite{petzsudar}.
That is, geodesic distances (as well as relative entropies) 
between density matrices {\it decrease} under 
coarse-grainings (completely positive trace-preserving maps, 
satisfying the Schwarz inequality: $T(a^{*}a) \geq T(a)^{*}T(a)$).
These metrics can be used for purposes of statistical 
distinguishability  \cite{petzsudar}.  The
monotone 
metrics for the TLQS 
have been found to be 
rotationally invariant over the Bloch ball, 
depending only on the radial coordinate 
$r$, that is the distance of the state in question 
from the origin $(0,0,0) $ --- corresponding to the fully mixed 
state. They are splittable  into
radial and tangential
 components of the form \cite[eq. (3.17)]{petzsudar},
\begin{equation} \label{radtang}
ds_{monotone}^2=\frac{1}{1-r^2} dr^2 +\Big((1+r) f(\frac{1-r}{1+r})\Big)^{-1} dn^2.
\end{equation}
Here, using 
spherical coordinates ($r,\theta_{1},\theta_{2}$), one has
 $dn^2 =  r^2 d \theta_{1}^2 +r^2 \sin^2{\theta_{1}} d \theta_{2}^2$.
Further, 
 $f: \mathbb{R}^{+} 
\rightarrow \mathbb{R}^{+}$ is an operator monotone function such that
$f(t)=t f(t^{-1})$ for every $t>0$. 
(A function is operator monotone 
if the relation $0 \leq K \leq H$, meaning 
that $H-K$ is nonnegative definite, implies $0 \leq f(K) \leq f(H)$ 
 for any such matrices $K$ and $H$ of any order.)
The radial component is {\it independent}
 of the function $f$, and in the case of
the Bures (minimal monotone) metric (corresponding to the 
particular choice
$f_{Bures}(t) = \frac{1+t}{2}$), the tangential component is 
independent 
of $r$ \cite{mjwhall}.

In the
classical 
context of the Husimi distribution, 
there is not an infinitude, but rather a {\it single} distinguished 
(up to a constant multiple) 
monotone 
metric --- the {\it Fisher information} metric 
\cite{chentsov,campbell,frieden}. 
(The counterpart here to stochastic mappings --- which are 
the appropriate morphisms in the category of quantum state 
spaces --- are stochastic {\it matrices}
\cite{petzsudar}.)
The $ij$-entry
 of the Fisher information matrix (tensor) is the expected value
with respect to the probability distribution in question
 of the product of the {\it first} derivative of the logarithm
of the probability with respect to its $i$-th parameter times the
analogous first 
derivative with respect to its  $j$-th parameter. (Under certain
regularity conditions, the Fisher information matrix 
 is equal to the ``second derivative matrix
for the informational divergence (relative entropy)''
\cite[pp. 455-456]{clarkebarron}, \cite[p. 43]{clarkebarron2}.)
The volume element of the Fisher information 
metric can be considered --- in the framework of Bayesian theory --- as a prior
distribution (Jeffreys' prior 
\cite{clarkebarron,slaterJeff,Kwek}) 
over, for our purposes here, the Bloch ball of TLQS. 
\subsection{Fisher information metric for the Husimi distribution}
\begin{figure} 
\includegraphics{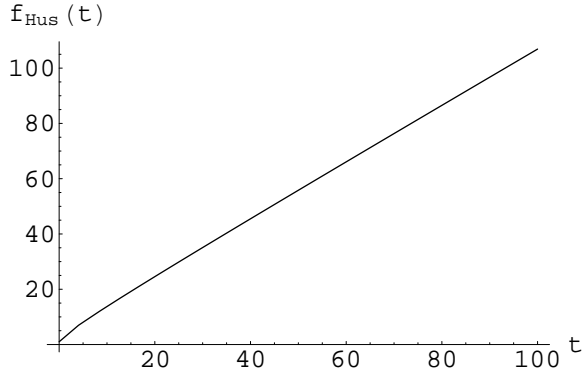}
\caption{\label{fig:graphtangential} The monotone function $f_{Hus}(t)$ 
that yields the {\it tangential} component of the Fisher information metric over the trivariate 
Husimi probability distributions for the two-level quantum systems.}
\end{figure}

We have found (having to make use of numerical, as well as 
symbolic MATHEMATICA 
procedures in our quest) 
that for the Husimi distribution over the TLQS, the Fisher 
information metric
takes the specific form (cf. (\ref{expressionHus})),
\begin{equation} \label{expressionHus}
ds^2_{Fisher_{Hus}} = \frac{-2 r - \log (\frac{1-r}{1+r})}{2 r^3} dr^2 +   
\Big((1+r) f_{Hus}(\frac{1-r}{1+r})\Big)^{-1} dn^2.
\end{equation}
Here,
\begin{equation} \label{zzz}
f_{Hus}(t)= \frac{(t-1)^3}{t^2-2 t \log{t}-1}.
\end{equation}
Now, a plot (Fig.~\ref{fig:graphtangential}) 
shows $f_{Hus}(t)$ to be, in fact,  a 
{\it monotone} function. ($f_{Hus}(t)$ is ``almost'' equal to $\frac{(t-1)^3}{t^2-2 t -1} =t-1$.) 
It has a singularity at $t=1$, corresponding to the fully mixed state ($r=0$),
 where $f_{Hus}(1+ \Delta t) \approx 3 + 3  \Delta t /2$,
though we have not attempted to confirm its {\it operator}
monotonicity. Also, $f_{Hus}(t)$ fulfills the self-adjointness condition
$f(t)=t f(t^{-1})$ of Petz and Sud\'ar \cite[p. 2667]{petzsudar}, at 
least at $t \neq 1$. For the pure states, that is $t=0,r=1$, we have
 $\lim_{t\to0} f_{Hus}(t)=1$.

We further  have the relation,
\begin{equation}
 c_{Hus}(p,q) = \frac{1}{q f_{Hus}(\frac{p}{q})} = \frac{q^2 -p^2 -2 p
q \log{\frac{q}{p}}}{(q-p)^3},
\end{equation}
 where $c_{Hus}(p,q)$ is a specific ``Morozova-Chentsov'' function. There exist one-to-one correspondences between Morozova-Chentsov functions, monotone metrics and 
operator means \cite[Cor. 6]{petzLIN}. ``Operator means are binary operations on positive operators which fulfill the main requirements of monotonicity and the transformer inequality'' \cite{petzLIN}.

\begin{figure}
\includegraphics{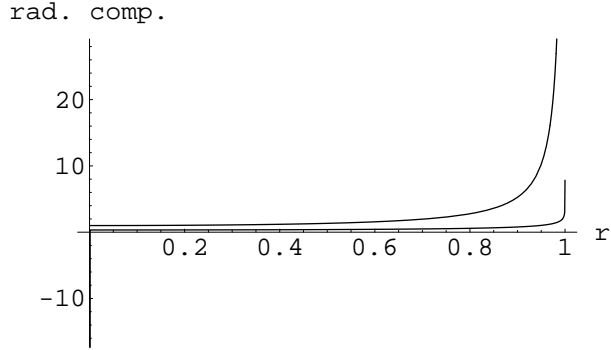}
\caption{\label{fig:2radials} The radial components of {\it any}
 monotone metric
and that of the Fisher information metric derived from the family of
trivariate
Husimi distributions over the TLQS. The one  for 
the (nondenumerably infinite) 
class $ds^2_{monotone}$ dominates that for $ds^2_{Fisher_{Hus}}$.}
\end{figure}

We can write (\ref{radtang}) more explicitly as
\begin{equation} \label{moreExplicit}
ds^2_{Fisher_{Hus}} = 
\frac{-2 r - \log (\frac{1-r}{1+r})}{2 r^3} dr^2 +   
\frac{ 2 r\ + (1-r^2) \log (\frac{1-r}{1+r})}{4 r^3} dn^2.
\end{equation}
Certainly,  $ds^2_{Fisher_{Hus}}$ does not have --- in 
terms of the radial 
component --- the specific 
form (\ref{radtang}) required of a monotone metric (cf. \cite{slaterlmp}).
In Fig.~\ref{fig:2radials} we show both  the {\it radial} components
of (any) $ds^2_{monotone}$ and of $ds^2_{Fisher_{Hus}}$.
Petz \cite[p. 934]{petzJPA} 
attributes the unvarying nature ($\frac{1}{1-r^2}$) of the radial component
of the (quantum) monotone metrics to the (classical) 
Chentsov uniqueness (of Fisher information) 
theorem \cite{chentsov,campbell}.
``Loosely speaking, the unicity [sic] result in the [probability]
simplex case survives along the diagonal and the off-diagonal provides
new possibilities for the definition of a stochastically invariant
metric'' \cite[p. 2664]{petzsudar}.

\begin{figure}
\includegraphics{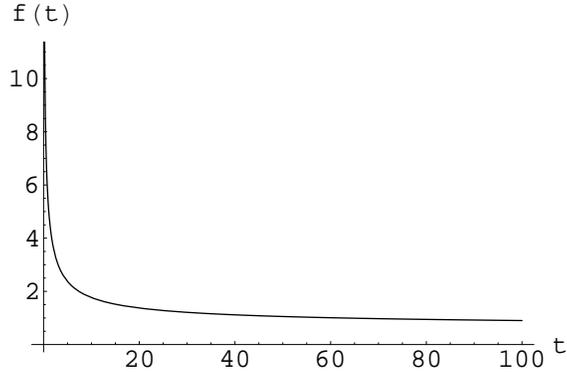}
\caption{\label{fig:Decreasing} Monotonically-{\it decreasing} function
$f_{counterfactual}$ 
obtained by equating the volume element of $ds^2_{Fisher_{Hus}}$ to that of
a generic monotone metric (\ref{radtang})}
\end{figure}

 If we 
(counterfactually) equate the volume element of
$ds^2_{Fisher_{Hus}}$ to that of a generic monotone metric
(\ref{radtang}), and 
solve for $f(t)$, we obtain a monotonically-{\it decreasing} function 
(Fig.~\ref{fig:Decreasing})
 (cf. \cite{slaterlmp}),
\begin{equation}
f_{counterfactual}(t)= 
\frac{{\sqrt{2}}\,{\left( -1 + t \right) }^{\frac{9}{2}}}
  {t\,\left( 1 + t \right) \,
    {\sqrt{{\left( -1 + t^2 - 2\,t\,\log (t) \right) }^2\,
        \left( 2 - 2\,t + 
          \left( 1 + t \right) \,\log (t) \right) }}}.
\end{equation}
Converting to cartesian coordinates $(x,y,z)$, 
the {\it trace} of $ds^2_{Fisher_{Hus}}$
can be simply expressed as $-\log{(\frac{1-R}{1+R})}/(2 R)$, where 
$R=\sqrt{x^2+y^2+z^2}$ (cf. \cite{luo,frieden}).
Also, at the fully mixed state ($x=y=z=0$), the metric is simply {\it flat}, that is
\begin{equation} \label{Flat}
ds^2_{Fisher_{Hus}} = \frac{1}{3} (dx^2 +dy^2 +dz^2).
\end{equation}
(The Riemann and Ricci tensors evaluated  at the fully mixed state 
 have no non-zero entries.)

Numerical evidence indicates that the 
Fisher information matrix for the 
Husimi distribution over the TLQS is 
bounded by the corresponding information 
matrices for the (quantum) monotone metrics, in the sense
that the monotone metric 
tensors minus the Fisher-Husimi information tensor are positive definite.

We can normalize the volume element of  $ds^2_{Fisher_{Hus}}$ 
to a probability distribution $p_{Hus}$ 
by dividing by the Fisher information metric {\it volume}  
$\approx 1.39350989367660$.
If we generate a ``hybridized-Husimi'' (quantum \cite{petzsudar})
monotone metric, $ds^2_{HYB_{Hus}}$, {\it via} the formula (\ref{radtang}), 
using  $f_{Hus}(t)$, then the volume of the
Bloch ball of TLQS in terms of this  newly-generated monotone
metric is $\frac{1}{2} \pi^2 (4 -\pi)
\approx 4.23607 > 1.39351$. Using this as a normalization factor, we obtain
a probability distribution ($p_{HYB_{Hus}}$) 
of interest over the TLQS.
\section{Comparative Noninformativities} \label{seccompnon}
Let us compare $p_{Hus}$  --- in the manner 
employed in \cite{compnoninform,clarke} 
(cf. \cite[sec. VI]{srednicki,slatercanosa}) --- with the 
prior probability distribution ($p_{Bures}$). The latter is gotten
by normalizing the volume element of the 
well-studied 
{\it minimal}  monotone (Bures) metric (\cite[eq. (7)]{slaterSPIN} \cite[eq. (16)]{slaterSPIN2}), that is, 
\begin{equation} \label{probBures}
p_{Bures} = \frac{r^2 \sin{\theta_{1}}}{\pi^2 \sqrt{1-r^2}},
\end{equation}
generated from (\ref{radtang}) 
using the operator monotone function $f_{Bures}(t)= \frac{1+t}{2}$.
(We avoid the 
specific 
designations $f_{min}(t)$ and $f_{max}(t)$ because these are usually, confusingly, considered to generate the maximal and minimal monotone metrics, respectively \cite[eq. (3.21)]{petzsudar}. 
Our integrations of probability distributions are conducted 
over $r \in [0,1], \theta_{1} \in [0,\pi]$ and $\theta_{2} \in [0,2 \pi]$.)

The {\it relative entropy} (Kullback-Leibler distance) of $p_{Bures}$ with respect to $p_{Hus}$ [which we denote $S_{KL}(p_{Bures},p_{Hus})$] --- that is, 
the {\it expected} value with respect to $p_{Bures}$
of $\log{\frac{p_{Bures}}{p_{Hus}}}$ --- is 
0.130845 ``nats'' of information. (We use the {\it natural} logarithm, and 
not 2 as a 
base, with one nat equalling 0.531 bits.) Let us note that the Shannon entropy
($S_{Shannon}$) 
of the Husimi distribution is the Wehrl entropy ($S_{Wehrl}$) 
of the corresponding
quantum state. Explicitly implementing \cite[eq. (6)]{mintert},
we have for the TLQS,
\begin{equation}
S_{Wehrl} =\frac{1}{4 r} 
\Big( 2 r + 4 r \log{2} +(1+r^2) 
\log{(\frac{1-r}{1+r})} -2 r \log{(1-r^2)} \Big).
\end{equation}
$S_{Wehrl}$ is always greater than the von Neumann
entropy, $S_{vN}= -\mbox{Tr}{\rho \ln{\rho}}$, which for the TLQS 
is expressible as
\begin{equation}
S_{vN}= \frac{1}{2} \Big( 2 \log{2} + r \log{(\frac{1-r}{1+r})} -\log{(1-r^2)} 
\Big).
\end{equation}
(We, of course, notice the omnipresence in these last two formulas, as well as in 
(\ref{moreExplicit}) and further formulas below  of
the term $W \equiv \log{(\frac{1-r}{1+r})}$. The two eigenvalues 
[$\lambda_{1}, \lambda_{2}=1-\lambda_{1}$] of $\rho$ are
$\frac{1 \pm r}{2}$, so $W$ is expressible as 
$\log{(\frac{\lambda_{1}}{\lambda_{2}})}$.)
Each monotone metric can be obtained in the form of a ``contrast functional''
for a certain convex subset of relative entropies \cite{lesniewski,jencova}.
\subsection{Bures prior} \label{Buresprior}
Now, let us convert $p_{Bures}$ 
to a {\it posterior} probability distribution ($post_{Bures}$) 
by assuming the performance of {\it six} measurements, {\it 
two} (with one outcome ``up'' and the other ``down'') 
in each of the $x-, y$- and $z$-directions. Normalizing 
the product of the prior $p_{Bures}$ and the {\it likelihood} 
function corresponding to the six measurement outcomes \cite[p. 3]{compnoninform},
\begin{equation} \label{jer}
post_{Bures} = \frac{192 p_{Bures} (1-x^2) (1-y^2) (1-z^2)}{71},
\end{equation}
 we find  $S_{KL}(post_{Bures},p_{Hus}) =  0.0912313 < 0.130845$. 
(The cartesian coordinates in (\ref{jer}) 
are transformed
to the spherical ones employed in our analysis.) So, in this sense 
 $p_{Bures}$ is {\it more} noninformative than $p_{Hus}$, 
the relative entropy being {\it reduced} by {\it adding} information
to $p_{Bures}$. 
On the other hand, $p_{Bures}$ --- corresponding to the {\it minimal}
 monotone metric --- is 
itself the {\it least}
 noninformative of the monotone-metric priors 
($p_{monotone}$) 
\cite{compnoninform}. 
(Luo has established an inequality between the [monotone metric] Wigner-Yanase
{\it skew information} and its minimal monotone counterpart \cite{luoWY}.)

Reversing the arguments of the relative entropy functional, we obtain 
$S_{KL}(p_{Hus},p_{Bures}) = .0818197$. 
But now, following the same form of posterior
construction, we find  $S_{KL}(post_{Hus},p_{Bures}) =0.290405 >.0818197$, 
further supportive of the
conclusion 
that $p_{Bures}$ is {\it more} 
noninformative than $p_{Hus}$. 
In some sense, then, $p_{Bures}$ assumes {\it less} about the data 
than $p_{Hus}$.
But this  diminishability of the relative entropy
is limited.  
If we convert $p_{Bures}$ to 
a new posterior $Post_{Bures}$ 
using the {\it square} of the 
likelihood function above --- that is, 
assuming {\it twelve} measurements, {\it four}  (with 
two  outcomes ``up'' and the other two ``down'')
in each of the $x-, y$- and $z$-directions, giving
\begin{equation} \label{PostBures}
Post_{Bures} = \frac{21504  p_{Bures} [(1-x^2) (1-y^2)  (1-z^2)]^m}{3793},
m=2,
\end{equation}
then $S_{KL}(Post_{Bures},p_{Hus}) = 0.292596 > 0.130845$.  
To much the same effect, 
if we  use a likelihood
 based on the {\it optimal/nonseparable}
set of measurements for {\it two} qubits,
consisting of five possible measurement outcomes, given in 
\cite[eq. (8)]{slaterefficiency}, to convert $p_{Bures}$ to a new posterior, 
then the relative entropy reaches 
 higher still, that is from 0.130845 to
0.623855. (Employing a likelihood
 based on the optimal/nonseparable
set of measurements for {\it three} qubits, 
consisting of eight possible measurement outcomes \cite[eq. (9)]{slaterefficiency}, the relative entropy with respect to $p_{Hus}$ increases further 
to 1.51365.) Actually, if we {\it formally} take 
$m=\frac{1}{2}$ in eq. (\ref{PostBures}), 
and renormalize to a new posterior,
we obtain a superior reduction, that is, to $0.07167 < 0.0912313$.
(Further, 
with $m=\frac{5}{8}$, we get 0.0702389 and  0.0732039, with $m=\frac{3}{4}$.)
\subsection{Morozova-Chentsov prior}
In \cite{compnoninform}, it was found that the (``Morozova-Chentsov'')
prior distribution,
\begin{equation} \label{probMC}
p_{MC} = \frac{.00513299 [ \log {\Big( \frac{1-r}{1+r} \Big)}]^2
\sin{\theta_{1}}}{\sqrt{1-r^2}},
\end{equation}
that is, the normalized volume element of the monotone metric (\ref{radtang})
based on the operator monotone function,
\begin{equation}
f_{MC}(t) = \frac{2 (t-1)^2}{(1+t) ({\log{t}})^2},
\end{equation} 
was apparently the {\it most}
noninformative of those (normalizable) 
priors based on the operator monotone functions
that had been explicitly 
discussed in the literature. Now, $S_{KL}(p_{MC},p_{Hus})= 1.37991$, that is, 
quite large.
This can be reduced to 0.893996 
if,  into $p_{MC}$,  
one incorporates $m=6$  measurements of the type described above; 
diminished further to 0.561901 
with $m=12$; and further still to 0.471852 --- the greatest 
reduction of this type --- with $m=18$. 
(For $m=24$, it starts to rise to 0.652441.) 

But, if we again
 use the likelihood 
based on the optimal nonseparable 
measurement of two qubits 
\cite[eq. (8)]{compnoninform}, with just five measurements, 
the relative entropy of the corresponding
posterior form of $p_{MC}$ with respect to $p_{Hus}$ 
is  reduced to 0.342124, which  is the {\it smallest}
 we have achieved so far
along these lines.
(For the mentioned optimal 
nonseparable measurement scheme 
for {\it three} qubits, the reduction is quite minor, only to 1.33492 nats.)
We obtained intermediate-sized 
reductions to 0.45524 and 0.492979, respectively, by using for our
measurements,
{\it twenty} projectors oriented to the vertices \cite[secs. 9, 10]{decker} 
of a dodecahedron and of an 
icosahedron. (The primary measurement scheme used above, and in \cite{compnoninform},
with six measurements oriented along three orthogonal directions, is 
tantamount to the use of an {\it octahedron}.)
\subsection{Hilbert-Schmidt prior}
The prior distribution generated by normalizing the volume element of the
Hilbert-Schmidt metric over the Bloch sphere is \cite[eq. (10)]{compnoninform}
\cite[eq. (31)]{mjwhall}
\begin{equation}
p_{HS} =3 \frac{r^2 \sin{\theta_{1}}}{4 \pi},
\end{equation}
which is simply the uniform distribution over the unit ball.
The Hilbert-Schmidt volume element can be reproduced using the formula 
(\ref{radtang})
for a quantum monotone metric, making use of $f_{HS}= \frac{(1+t)^2}{\sqrt{t}}$, but this function is neither monotone-increasing nor decreasing over
$t \in [0,1]$ (cf. \cite{ozawa}).

We have that $S_{KL}(p_{Hus},p_{HS}) = 0.0579239$ and $S_{KL}(p_{HS},p_{Hus})
= 0.05443$. Now, in terms of our usual posterior distributions based
on six measurements, $S_{KL}(post_{Hus},p_{HS}) = 0.0236596$ and
$S_{KL}(post_{HS},p_{Hus}) = 0.278953$, so we can conclude that 
the Husimi prior $p_{Hus}$
is more noninformative than the Hilbert-Schmidt prior $p_{HS}$.
\section{Universal Data Compression}
Employing $p_{Hus}$ 
as a prior distribution (Jeffreys' prior) 
over the family (Riemannian
manifold) of Husimi qubit probability distributions, 
 the (classical) {\it asymptotic minimax/maximin
 redundancy}
of {\it universal data compression} is equal
to  \cite[eq. (2.4)]{clarkebarron2} \cite{clarkebarron},
\begin{equation} \label{classicalresult}
\frac{3}{2} \log{\frac{n}{2 \pi e}} + \log 1.39350989367660 =\frac{3}{2} \log{\frac{n}{2 \pi e}} +0.331826 = \frac{3}{2} \log{n} -3.92499,
\end{equation}
where $n$ is the sample size (the number of qubits [TLQS]) 
and we used the before-mentioned volume of $ds^2_{Fisher_{Hus}}$.
(``Suppose that $X$ is a discrete random variable whose distribution is in the 
parametric family $\{P_{\theta}:\theta \in \Theta \}$ and we want to encode a 
block of data for transmission. It is known that a lower bound on the expected codeword length is the entropy of the distribution. Moreover, this entropy bound can be achieved, within one bit, when the distribution is known.
Universal codes have expected length near the entropy no matter which member
of the parametric family is true. The redundancy of a code is defined to be
the difference between its expected length and its entropy'' \cite[p. 459]{clarkebarron}.)

For the {\it quantum}/nonclassical 
 counterpart \cite{krattenthaler} (cf. \cite{hayashi,jozsa,jozsa1}), let us consider 
the use of the
``Grosse-Krattenthaler-Slater'' 
(``quasi-Bures'') probability distribution  \cite[eq. (33)]{slaterefficiency},
\begin{equation}
p_{GKS}= \frac{0.0832258 e}{1-r^2} \Big(\frac{1-r}{1+r} \Big)^{\frac{1}{2 r}}
 r^2
\sin{\theta_{1}}.
\end{equation}
This is the normalized form of the monotone metric (\ref{radtang})
associated with
the (presumably operator) monotone function,
\begin{equation}
f_{GKS}(t) =\frac{t^{t/(t-1)}}{e}.
\end{equation}
(Taking limits, we have for the fully mixed state, $f_{GKS}(1)=1$ and 
for the pure states, $f_{GKS}(0)= e^{-1}$.)
It appears \cite{asyred2}
(though not yet fully rigorously established)
that the (quantum) asymptotic minimax/maximin  redundancy, employing
$p_{GKS}$ as a  
prior probability distribution over the $2 \times 2$ 
density matrices ({\it and} their $n$-fold 
tensor products (cf. \cite{caves})), is
$\frac{3}{2} \log{n} -1.77062$. This is
{\it greater} than the classical (Husimi-Fisher-information-based) analog
(\ref{classicalresult}) by 2.20095 nats of information.
It would seem that 
this difference is attributable to the greater dimensionality
($2^n$) of an $n$-qubit Hilbert space, as opposed to a dimensionality
of $3 n$ for $n$ trivariate 
Husimi probability distributions over the TLQS.

\begin{figure}
\includegraphics{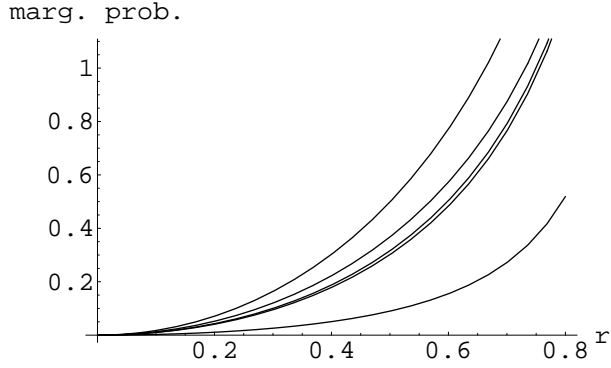}
\caption{\label{fig:BuresHYBRID}Plots of one-dimensional marginal probability distributions over the radial coordinate $r$ of $p_{Bures}$, 
$p_{HYB_{Hus}}$, $p_{GKS}$, $p_{Hus}$ and $p_{MC}$. The order of dominance 
of the curves is:
$p_{Hus}>p_{Bures}>p_{GKS}>p_{HYB_{Hus}}>p_{MC}$. The 
marginal distributions of $p_{HYB_{Hus}}$ and $p_{GKS}$ are quite close, as reflected in their small relative entropy  ($\approx .0004$).}
\end{figure}

We further note that $S_{KL}(p_{Bures},p_{HYB_{Hus}}) =0.00636046$
and $S_{KL}(p_{HYB_{Hus}},p_{Bures}) = 0.0062714$, both being very small.
Smaller still, 
$S_{KL}(p_{Bures},p_{GKS})= 0.00359093$ and $S_{KL}(p_{GKS},p_{Bures}) = 0.00354579$ --- whence the designation $p_{quasi-Bures} \equiv p_{GKS}$. 
But then, even more strikingly, we computed that
$S_{KL}(p_{GKS},p_{HYB_{Hus}})= 0.000397852$ and $S_{KL}(p_{HYB_{Hus}},p_{GKS})= 
0.000396915$.
In Fig.~\ref{fig:BuresHYBRID} we show the {\it one}-dimensional marginal probability distributions over the radial coordinate $r$ of the five 
distributions $p_{Bures}$,
$p_{HYB_{Hus}}$, $p_{Hus}$, $p_{GKS}$ and $p_{MC}$, with those for
$p_{HYB_{Hus}}$ and  $p_{GKS}$ being --- as 
indicated --- particularly proximate.

Substitution of $p_{HYB_{Hus}}$ for $p_{GKS}$ 
into the quantum asymptotic (maximin) redundancy formula that has to
be {\it maximized} over all possible prior probability
distributions \cite[eq. (4.3)]{asyred2},
\begin{multline}\label{c2} 
\frac {3} {2}\log n  -\frac {1} {2}- \frac {3} {2} \log 2  
- \frac {3} {2} \log \pi \\
+4\pi\int _{0} ^{1}\left(-\log(1-r^2)
+\frac {1} {2r}\log\left(\frac {1-r} {1+r}\right)
-\log w(r)\right)r^2w(r)\,dr,
\end{multline}
leads to a very slightly decreased (and hence suboptimal) 
redundancy, $\frac{3}{2} \log{n} - 1.77101$ 
{\it vs.} $\frac{3}{2} \log{n} - 1.77062$. (Use of $p_{Bures}$ 
as a quantum prior over the $2 \times 2$ density matrices gives us a 
constant term of $-1.77421$, use 
of $p_{Hus}$, $-1.88279$ and use of $p_{MC}$, $-2.15667$.)
To obtain the appropriate form of $w(r)$ to use in (\ref{c2}), we take our
probability distributions (such as (\ref{probBures}) and (\ref{probMC})), 
divide them by $4 \pi r^2$ and integrate the results
over $\theta_{1} \in [0,\pi]$ and $\theta_{2} \in [0,2 \pi]$.
 (Thus, we must
have $4 \pi \int_{0}^{1} w(r) r^2 dr =1$.)
The {\it minimax} objective function is 
\begin{equation}\label{d6}
\min_{w}\max_{0\le r\le 1}
\bigg(\frac {3} {2}\log n  -\frac {1} {2}- \frac {3} {2} \log 2  
- \frac {3} {2} \log \pi -\log(1-r^2)
+\frac {1} {2r}\log\left(\frac {1-r} {1+r}\right)
-\log w(r)\bigg).
\end{equation}
The minimax is also achieved using the $w(r)$ formed from $p_{GKS}$.

We can, additionally, 
achieve an extremely good fit to $p_{Hus}$ by proceeding in 
somewhat an {\it opposite} fashion to that above --- {\it reversing} our
hybridization 
procedure. Employing $f_{GKS}(t)$, rather than
$f_{Hus}(t)$ in the expression (\ref{expressionHus}) 
for $ds^2_{Fisher_{Hus}}$ and obtaining the 
corresponding normalized (dividing by 4.00277) 
volume element ($p_{\tilde{HYB}_{GKS}}$), we find
$S_{KL}(p_{\tilde{HYB}_{GKS}},p_{Hus}) 
= 0.000316927$. (Interchanging the arguments of
the relative entropy functional, we get 0.000317754.)
It is quite surprising, then, that a joint plot
of $f_{GKS}(t)$ and $f_{Hus}(t)$ 
readily shows them to be substantially 
{\it different}
in character (for example, $f_{Hus}(50)=55.8161$ and $f_{GKS}(50)= 19.9227$),
since they have been shown here to generate two pairs of such highly
similar probability distributions, 
one pair composed of 
(quantum) 
monotone ($p_{GKS}$ 
and $p_{HYB_{Hus}}$), and the other pair of
(quantum) non-monotone metrics ($p_{\tilde{HYB}_{GKS}}$ and $p_{Hus}$).

\section{Escort-Husimi Distributions} \label{EscortSec}
For the {\it escort}-Husimi distributions \cite{pennini}, we raise the probability element of the Husimi distribution to the $q$-th power, and renormalize to
a new probability distribution. (Of course, the Husimi distribution itself
corresponds to $q=1$. If we set $\alpha=2 q -1$, we recover the 
$\alpha$-family of Amari \cite{amari,naudts2,jencova}.) 
To normalize the $q$-th power of the Husimi distribution, one must divide by
\begin{equation}
\frac{2^{-q} \Big(-(1-r)^{1+q} +(1+r)^{1+q} \Big)}{r +q r}.
\end{equation}
\subsection{The case $q=2$}
For (entropic index) 
$q=2$, the 
Fisher information metric takes the form
\begin{equation} \label{q2}
ds^2_{Fisher_{q=2}} = \frac{12}{(3+r^2)^2} dr^2 + \Big( (1+r) 
f_{q=2}(\frac{1-r}{1+r}) \Big) dn^2,
\end{equation}
where
\begin{equation}
f_{q=2}(t) = \frac{t^2+t+1}{2(t+1)}.
\end{equation}
We have $f_{q=2}(1) = \frac{3}{4}$ and $f_{q=2}(0)=\frac{1}{2}$.

\subsubsection{Relative entropies}
Further, the relative entropies $S_{KL}(p_{Hus},p_{Esc_{q=2}}) =0.0114308$ and
$S_{KL}(p_{Bures},p_{Esc_{q=2}}) =0.42964$, So, it appears that $p_{Esc_{q=2}}$ is even less
noninformative than $p_{Hus}$  (recalling that $S_{KL}(p_{Bures},p_{Hus}) = 0.130845 < 0.42964$), which in turn we found above was less noninformative
than the prior probabilities formed from any of the 
(quantum) monotone metrics.
We also note that $S_{KL}(post_{Bures},p_{Esc_{q=2}}) = 0.125159 <0.42964$.
If we ``hybridize''
$ds^2_{Fisher_{q=2}}$ by modifying its radial
component into that required of a (quantum) monotone metric, 
then we find that $S_{KL}(p_{Bures},p_{HYB_{q=2}}) = 0.00246031
(< (S_{KL}(p_{Bures},p_{HYB_{Hus}}) = 0.00636046)$ is quite small.
\subsection{The cases $q>2$}
 For the escort-Husimi probability distribution with 
$q=3$, the Fisher information metric takes the
 form
\begin{equation}
ds^2_{Fisher_{q=3}}= \frac{3-r^2}{(1+r^2)^2} dr^2 + \Big( (1+r)
f_{q=3}(\frac{1-r}{1+r}) \Big) dn^2,
\end{equation}
where
\begin{equation}
f_{q=3}(t) = \frac{t^2+1}{3(t+1)}.
\end{equation}
Now, $f_{q=3}(1) =f_{q=3}(0) = \frac{1}{3}$ and a plot of $f_{q=3}(t)$ 
clearly manifests  monotonic behavior also. (The monotonically-decreasing
scalar curvature of $ds^2_{Fisher_{q=3}}$ equals $\frac{4}{3}$ at $r=0$.)
We have that $S_{KL}(p_{Bures},p_{Esc_{q=3}}) = 0.63705 > S_{KL}(p_{Bures},p_{Esc_{q=2}}) = 
0.42964$, so the informativity (noninformativity) 
of the escort-Husimi prior probabilities
{\it appears} to increase (decrease) with $q$.

For $q=4$,
\begin{equation}
ds^2_{Fisher_{q=4}} = \frac{80 (5-2 r^2 +r^4)}{3 (5 + 10 r^2 +r^4)^2} dr^2
+ \Big( (1+r) f_{q=4}(\frac{1-r}{1+r}) \Big)^{-1} dn^2,
\end{equation}
where
\begin{equation}
f_{q=4}(t) =\frac{3 (t^4 +t^3+t^2+t+1)}{4 (t+1) (3 t^2 + 4 t+ 3)}.
\end{equation}

For $q=5$,
\begin{equation}
ds^2_{Fisher_{q=5}} = \frac{3 (5-r^2) (5+3 r^4)}{(3 +10 r^2 + 3 r^4)^2} dr^2 + 
\Big( (1+r) f_{q=5}(\frac{1-r}{1+r}) \Big)^{-1} dn^2,
\end{equation}
where
\begin{equation} 
f_{q=5}(t) = \frac{2 (t^4 +t^2 +1)}{5 (t +1) (2 t^2 +t +2)}.
\end{equation}
We have (as found  by C. Krattenthaler, making use of 
explicit MATHEMATICA computations of ours 
for $q=2, 3,\cdots,40$) (cf. \cite[sec. 3.2]{gnutzmann} \cite{sugita}),
\begin{equation} \label{ck}
f_{q}(t) = \frac{(q-1)  \Sigma_{i=0}^{q} t^i}{q (t+1) \Sigma_{i=1}^{q-1} 
i (q-i)  t^{i-1}}.
\end{equation}
(For {\it odd} $q$ some simplification 
in the resulting expression occurs due to cancellation by a factor
of $(t+1)$.)

\begin{figure}
\includegraphics{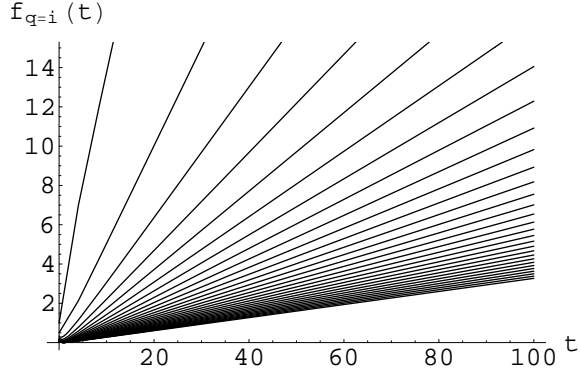}
\caption{\label{fig:Escortq=i}The monotone functions $f_{q=i}(t), i=
1,\cdots,30$ that yield
the {\it tangential} components of the Fisher information metric over the
escort-Husimi ($q=i$) probability distributions. The steepness of the 
graphs decreases as $q$ increases}
\end{figure}

In Fig.~\ref{fig:Escortq=i} we plot $f_{q=i}(t), i = 1,\cdots,30$, 
revealing their common monotonically-increasing behavior. (Of course, 
we have $f_{q=1}(t)
\equiv f_{Hus}(t)$, shown already in Fig.~\ref{fig:graphtangential}. 
The steepness of the curves {\it decreases} with
increasing $q$.)

Let us further note that in addition to 
$S_{KL}(p_{Bures},p_{HYB_{Hus}}) = 0.00636046$ and
$S_{KL}(p_{Bures},p_{HYB_{q=2}}) = 0.00246043$,
we have $S_{KL}(p_{Bures},p_{HYB_{q=3}}) = 0.0132258, S_{KL}(p_{Bures},p_{HYB_{q=4}}) = 
0.0238858$ and $S_{KL}(p_{Bures},p_{HYB_{q=5}}) = 0.0327578$. 
(We have also been able to compute that $S_{KL}(p_{Bures},p_{HYB_{q=1000}}) = 
0.0969315$ and $S_{KL}(p_{GKS},p_{HYB_{q=1000}})= 0.127027$.) So, the best of these fits of  $p_{Bures}$ to the prior probabilities for the hybridized-escort-Husimi
probability distributions is for $q=2$.
\subsection{Tangential components}
Now, we can reexpress the formula (\ref{ck}) {\it without} summations, 
making use of the binomial theorem, as
\begin{equation} \label{nonintegral}
f_{q}(t) = \frac{\left( -1 + q \right) \,{\left( -1 + t \right) }^2\,
    \left( -1 + t^{1 + q} \right) }{q\,
    \left( 1 + t \right) \,
    \left( 1 - q + t + q\,t - t^q - q\,t^q - t^{1 + q} + 
      q\,t^{1 + q} \right) }.
\end{equation}
So, we could study hybridized escort-Husimi metrics based on 
{\it non}-integral $q$ using this formula.  (We note that 
(\ref{nonintegral}), in fact, yields
$\lim_{q \to 1}f_{q}(t) \equiv f_{Hus}(t)$.)
For example,
\begin{equation}
f_{q=\frac{1}{2}}(t)= 6 + 6 \sqrt{t} +2 t -\frac{4}{1+t}.
\end{equation}
Thus, (\ref{nonintegral}) gives us (following the formulation (\ref{radtang}))
the tangential components of the escort-Husimi Fisher information metrics
for arbitrary $q$. (Pennini and Plastino \cite{pennini} 
have argued, though,  that in a {\it quantal} 
regime, $q$ can be no less than 1. Tsallis statistics with 
an entropic index of $q=\frac{3}{2}$, Beck has contended, correctly describes
the small-scale statistics of Lagrangian turbulence \cite{beck}.) 
\subsection{Radial components}
We do not have,  at this point, a comparable 
complete 
formula for the {\it radial} components.
However, C. Krattenthaler has shown --- making use of explicit computations of
ours for the cases $q=2,3,\cdots,18$ --- that 
the {\it denominators} of the functions
giving the radial components are simply proportional to
\begin{equation} \label{Pochhammer}
u(q) = \Big(\Sigma_{i=0}^{q}[\frac{Pochhammer[q - 2 i + 1, 2 i + 1] r^{2 i}}{2 (2 i + 1)!}
\Big)^{2}.
\end{equation}
(The Pochhammer symbol is synonymous with the rising or ascending factorial.
The  obtaining of  comparable formulas for the {\it numerators} 
of the radial components 
 might be possible using the 
``Rate.m'' program available from the website of Krattenthaler 
[http://www.mat.univie.ac.at/~kratt/], 
if we had available additional 
explicit computations beyond the $q=18$.)
As way of illustration, the radial component of
 $ds^2_{Fisher_{q=8}}$ is expressible as
\begin{equation}
\frac{144\,\left( 21 + 42\,r^2 + 135\,r^4 + 28\,r^6 + 
      35\,r^8 - 6\,r^{10} + r^{12} \right) }{7 u(8)}.
\end{equation}

\section{Positive P-Representation for TLQS} \label{PositiveSec}

Braunstein, Caves and Milburn focused on a specific choice of
{\it positive} 
P-representation which they called the canonical form and which is
always well defined \cite[eq. (3.3)]{bmc} (cf. \cite[sec. 6.4]{crispin}):
\begin{equation} \label{bmcEQ}
P_{can}(\alpha,\beta^*) \equiv \frac{1}{4 \pi^2} \exp{(-\frac{1}{4} |\alpha-\beta|^2)}<\frac{1}{2} (\alpha+\beta)|\hat{\rho}|\frac{1}{2} (\alpha+\beta)>
=\frac{1}{4 \pi^2} \exp{(-\frac{1}{4}|\alpha-\beta|^2)} Q(\frac{1}{2}
(\alpha+\beta)).
\end{equation}
``The canonical form is clearly positive, and...it is essentially the Q-function [Husimi distribution]'' \cite{bmc}.

We sought to implement this model, choosing for $\alpha$ and $\beta$ 
{\it independent} 2-dimensional representations of the spin-$\frac{1}{2}$ 
 coherent states 
(while for the Husimi distribution or Q-function, only, say $\alpha$, need be
employed).
(The ``positive P representation achieves [its] considerable success by
doubling the number of degrees of freedom of the system, i. e., 
doubling the number of dimensions of the phase space'' \cite[p. 1153]{bmc}.
More typically, in the positive P-representation, 
$\alpha$ and $\beta$ are allowed to 
vary independently over the {\it entire} complex
plane.)
However, then our result --- using this choice of $\alpha$ and 
$\beta$ --- was {\it not} normalized to a probability distribution
in the manner indicated in (\ref{bmcEQ}). 

We noted that Braunstein,
Caves and Milburn had commented that a ``positive P
representation can be defined for a large class of operators. We restrict
ourselves here to those that are built up from the standard annihilation
and creation operators of a harmonic oscillator. In particular, our work does
not apply to generalizations of the positive P representation that include
spin or pseudospin operators often used to describe a two-level atom''
 \cite[p. 1155]{bmc}.
(We are not aware, however, of any specific applications 
reported in the literature of the
positive P-representation to $n$-level [finite-dimensional] quantum systems.)

\begin{figure}
\includegraphics{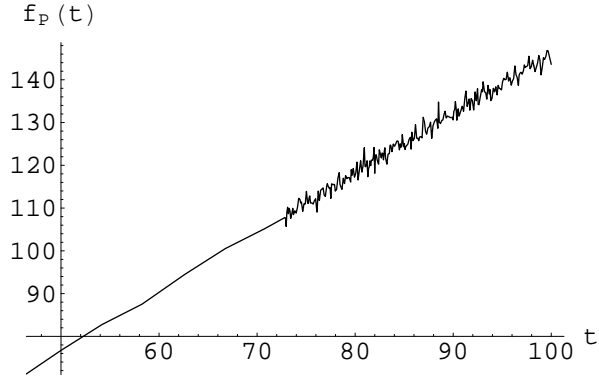}
\caption{\label{fig:positiveP3tang} {\it Approximation} to the presumed operator monotone function $f_{P}(t)$ yielding the {\it tangential} component of
$ds^2_{Fisher_{P}}$ for the positive P-representation over the two-level
quantum systems}
\end{figure}

\begin{figure}
\includegraphics{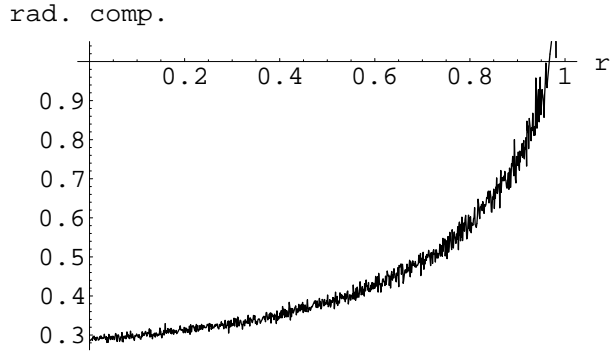}
\caption{\label{fig:positiveP3rad} {\it Approximation} to the {\it radial}
component of $ds^2_{Fisher_{P}}$ for the positive P-representation over the
two-level quantum systems}
\end{figure}

We did not perceive how to exactly 
(re)normalize the distribution (\ref{bmcEQ}) for 
our particular choices of $\alpha$ and $\beta$. So, we expanded just the 
exponential term of (\ref{bmcEQ}) 
into a power series in third order in the four {\it phase}
variables and exactly normalized  the product of this series with the 
remaining unmodified factor (the Q-function or Husimi distribution) to 
obtain a new (presumed) probability distribution.
We then fit (numerically) the resultant tangential component 
of the associated Fisher information metric to the form
(\ref{radtang})
required of a monotone metric. In Fig.~\ref{fig:positiveP3tang} we show what
we (gratifyingly)  obtained in this manner for 
$f_{P}(t)$.
In Fig.~\ref{fig:positiveP3rad} we show an approximation to the radial
component of $ds^2_{Fisher_{P}}$, similarly obtained.
(The positive P-function ``seems to possess some interesting properties and may
deserve close inspection''\cite[p. 175]{lee}.)
It would be of interest to see how near 
the associated probability 
distributions ($p_{P}$ 
and $p_{HYB_{P}}$) would be to the probability distributions
(already discussed above) $p_{GKS},p_{Hus},p_{HYB_{Hus}}$ and
$p_{\tilde{HYB}_{GKS}}$. Most pressing, though, is the question of whether
or not the concept of a positive P-representation does, in fact, 
have  a meaningful and
natural theoretical application to the $n$-level quantum systems.

\section{Gaussian Distribution} \label{GaussianSec}

\begin{figure}
\includegraphics{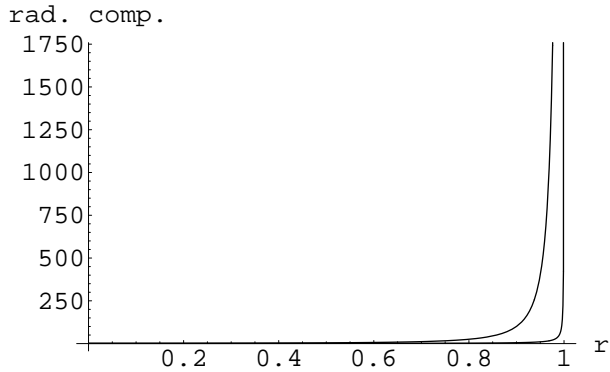}
\caption{\label{fig:QuasiMaximal}Radial
components of $ds^2_{monotone}$ 
and $ds^2_{Fisher_{Gauss}}$. The latter dominates the former.}
\end{figure}

An approach quite distinct from that of the Husimi probability distributions,
but still {\it classical} in nature, to modeling quantum systems has been presented
in \cite{guiasu,levine,bach,bach2,bach3} (cf. \cite{gibbons}). 
Here the family of 
probability distributions is taken as that of the Gaussian (complex multivariate normal distributions) having covariance matrix equal to the density matrix.
For the TLQS, Slater \cite[eq. (13)]{quantumcointossing} \cite[eq. (16)]{physicaA} derived the corresponding
Fisher information metric. This is representable as,
\begin{equation} \label{quasiMaximal}
ds^2_{Fisher_{Gauss}}
 =  \frac{2 (1+r^2)}{(1-r^2)^2} dr^2 + \frac{2}{1-r^2}dn^2.
\end{equation}
The tangential component can be reproduced, following the basic formula 
(\ref{radtang}), 
by choosing  $f_{Gauss}(t)=\frac{t}{1+t}$.
This is simply {\it one-half} of 
that --- $f_{YL}(t) = 2 f_{Gauss}(t) = 
 \frac{2 t}{1+t}$ --- associated with the {\it maximal} monotone (Yuen-Lax) 
metric \cite{YuenLax}. Like that metric,
the metric (\ref{quasiMaximal}) yields a {\it non}-normalizable volume element (so one can not immediately apply --- without some 
preliminary 
truncation --- the 
comparative noninformativity/relative entropy test we have used above 
\cite{compnoninform,clarke}).
Of course, the radial component of (\ref{quasiMaximal}) is 
also not consistent
with the requirement for a monotone metric. In fact, it rises 
much {\it more} steeply
than $\frac{1}{1-r^2}$, in opposite behavior 
 to that for $ds^2_{Fisher_{Hus}}$.
In Fig.~\ref{fig:QuasiMaximal} we show this phenomenon.

\section{Discrete Wigner Function for a Qubit} \label{WignerSec}
The discrete Wigner function (pseudoprobability) 
$W$, in the simplest case of a qubit, 
is defined
on a $2 \times 2$ array, 
with four components $W_{i,j},i, j =1,2$ \cite[eqs. (14)-(17)]{galvao}.
The sum of $W_{ij}$ in each ``line'' $\lambda$ is the probability 
$p_{ij}$ of projecting the state onto the basis vector $|\alpha_{ij}>$,
where $i \in \{1,2,3\}$ indexes a set of three mutually unbiased bases 
(MUB) for
a qubit and $j \in \{1,2\}$ indexes the basis vector in each MUB.
Choosing the MUB to be the eigenstates of the three Pauli operators,
and using our cartesian coordinates, one can obtain three one-dimensional
{\it marginal} (binomial) probability distributions over the $x-, y-$ and $z$-axes, 
of the form $(\frac{1+x}{2},\frac{1-x}{2}),\cdots$ 
(cf. \cite{asplund,cunha}). Now, the corresponding
Jeffreys' prior for the one-dimensional family of such 
 binomial distribution is simply the {\it beta} distribution
$p_{\beta}(x) = \frac{1}{\pi \sqrt{1-x^2}}$. (Let us note that the
one-dimensional marginal distributions obtained 
for $p_{Bures}$ are of another form,
that is, $\frac{2 \sqrt{1-x^2}}{\pi}$.)

Let us take the product of $p_{\beta}(x),p_{\beta}(y)$ and 
$p_{\beta}(z)$, which naturally forms a (prior) 
probability distribution,
\begin{equation}
p_{product} = \frac{1}{\pi^3 \sqrt{(1-x^2) (1-y^2) (1-z^2)}},
\end{equation}
over the {\it hypercube}
with vertices $(\pm 1,\pm 1,\pm 1)$
and renormalize/truncate
 it to a probability distribution over the Bloch sphere,
\begin{equation}
p_{Wigner} = \frac{1}{6.61455516101  \sqrt{(1-x^2) (1-y^2) (1-z^2)}}.
\end{equation}
(Thus, the quantum-mechanically {\it inaccessible} region lying outside the
Bloch ball, but within the hypercube is disregarded --- assigned null 
measure --- in the new
normalization.)

Now, we found --- strictly following the notation, formulas 
and line of argument above
in sec.~\ref{seccompnon} --- that 
$S_{KL}(p_{Wigner},p_{Hus}) = 0.0149831$ 
and $S_{KL}(p_{Hus},p_{Wigner}) =0.0156225$,
so these two distributions are rather close in nature. Of course,
$p_{Hus}$ is rotationally-symmetric over the Bloch sphere, while 
$p_{Wigner}$ is not, so it seems to make little sense to try to
compute some function $f_{Wigner}(t)$ to generate the tangential component.
We found it problematical, using our usual (relative entropy) approach,
to designate either $p_{Hus}$ or $p_{Wigner}$ as more or less noninformative.
(The ``Husimi function is a kind of...coarse-grained Wigner function''
\cite[p. 3]{sugita}.)

\section{Scalar curvature}
\begin{figure}
\includegraphics{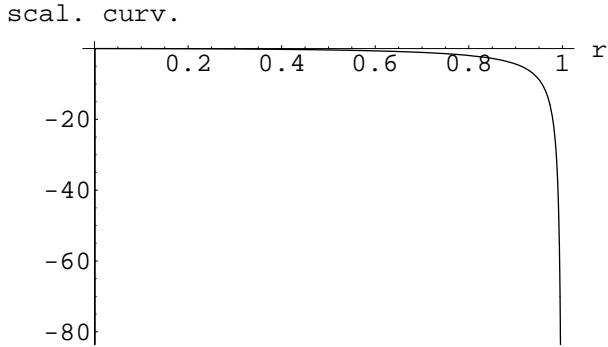}
\caption{\label{fig:HusScalCurv} Scalar curvature of the Fisher information
metric for the family of Husimi distributions}
\end{figure}

In Fig.~\ref{fig:HusScalCurv}, we plot the {\it scalar curvature} of
$ds^2_{Fisher_{Hus}}$.
The formula for this scalar curvature is
\begin{equation}
K^{n=2}_{Hus}= \frac{r\,\left( -6\,r + W\,\left( -3 + r^2 \right) 
      \right) \,\left( -4\,r^2\,
       \left( -3 + r^2 \right)  + 
      6\,W\,r\,\left( 2 - 3\,r^2 + r^4 \right)  + 
      W^2\,\left( 3 - 8\,r^2 + 5\,r^4 \right)  \right) }
    {{\left( W + 2\,r \right) }^2\,
    \left( -1 + r^2 \right) \,
    {\left( -2\,r + W\,\left( -1 + r^2 \right)  \right) }^
     2},
\end{equation}
where $W=\log{\frac{1-r}{1+r}}$.
Also, expanding about $r=0$,
\begin{equation}
K^{n=2}_{Hus} \approx \frac{-6\,r^2}{5} - \frac{138\,r^4}{125} - 
  \frac{32094\,r^6}{30625} - \frac{154474\,r^8}{153125} - 
  \frac{57710054\,r^{10}}{58953125}.
\end{equation}
The {\it nonpositive} monotonically-{\it decreasing} 
scalar curvature (Fig.~\ref{fig:HusScalCurv}) has its {\it maximum} at $r=0$,
corresponding to the fully mixed state, indicative of a {\it flat}
metric there (cf. (\ref{Flat})) (and is $-\infty$ at the pure states,
$r=1$).
For the minimal monotone (Bures) metric, the {\it nonnegative} scalar
curvature is {\it constant}, that is $K^{n=2}_{min} = 6$, over the Bloch ball, and for the 
$(n^2-1)$-dimensional convex set of $n \times n$ density matrices, $n>2$,
achieves its {\it minimum} of $K^{n}_{min}=\frac{(5 n^2-4)(n^2-1)}{8}$ 
at the fully mixed state ($\rho=\frac{1}{n} I$)
\cite{dittmannjgp}. (In \cite{dittmannjgp}, the metric used is one-quarter of
that corresponding to (\ref{radtang}), used here, so the results we compute here differ from those there by such a factor.
For the {\it maximal} monotone metric, $K^{n=2}_{max}=\frac{8 (r^2-6)}{1-r^2}$, which is monotonically-{\it decreasing} as $r$ increases, as is 
$K^{n=2}_{Hus}$.)

For the two-level quantum sytems, Andai \cite{attila} has constructed a 
family of monotone metrics with {\it non}-monotone scalar curvature, and
given a condition for a monotone metric to have a local minimum at the
maximally mixed state.

\subsection{Metrics of constant scalar curvature}
The metric $
ds^2_{Fisher_{q=2}}$ has {\it constant} scalar curvature, $K^{n=2}_{q=2} 
=\frac{3}{2}$, while, as previously noted, 
 $K^{n=2}_{min}=6$. Let us note that 
$K^{n}_{WY}=\frac{1}{4} (n^2-1) (n^2-2)$, which is also 
$\frac{3}{2}$ for $n=2$.
Here,  WY denotes the Wigner-Yanase metric --- the only pull-back metric
among the quantum monotone metrics --- and 
$f_{WY}(t)=\frac{1}{4} (\sqrt{t}+1)^2$, which is the only {\it self-dual} 
operator monotone function \cite{pablo}. 
``It is not known at the moment if there are other monotone metrics of 
constant sectional and scalar curvature'' \cite[p. 3760]{pablo}.
It is a theorem that the ``set of two-dimensional normalized density matrices
equipped with the Bures metric is isometric to one closed-half of the three-sphere with radius $\frac{1}{2}$'' \cite{hubner}. 
The WY-metric ``looks locally like a sphere of radius 2
of dimension $(n^2-1)$''
\cite[p. 3759]{pablo}.
 If we transform to spherical coordinates on the 
3-sphere, then, the metric tensor for $ds^2_{min}$ is diagonal in character,
while the two other (constant scalar curvature) metrics are not  (cf. \cite{lajos}).

The three metrics $ds^2_{min}, ds^2_{WY}$ and $ds^2_{Fisher_{q=2}}$ 
are {\it Einstein}.
If we {\it scale} these  metrics so that they are all of {\it unit} volume 
\cite{gursky},
then $K^{n=2}_{min/scaled} = 6 \pi^2 \approx 59.2176,
K^{n=2}_{WY/scaled}= 6 \pi (\pi -2) \approx 21.5185$ and
$K^{n=2}_{q=2/scaled}= 4 \pi^2 - 6 \sqrt{3} \pi \approx 6.83003$.
The constant scalar curvatures of (unit-volume) 
{\it Yamabe} metrics 
are bounded above, and their least upper bound is a real number equal to
$n (n-1) V_{n}^{2/n}$, where $V_{n}$ is the volume of the standard metric
on $S^{n}$, and in our (Bloch sphere) case, $n=3$, so the bound is 
$24 2^{\frac{1}{3}} \pi^{\frac{4}{3}} \approx 139.13$ \cite{gursky}.

\section{Discussion}

Luo \cite{luo} (cf. \cite[sec. 2.4]{pennini} \cite{twamley,slaterJeff,Kwek}) 
has 
calculated the 
Fisher information matrix of the Husimi distribution  in the 
Fock-Bargmann representation 
of the quantum harmonic oscillator with one degree of freedom.  
He found that the Fisher information of the position and that of the momentum move in opposite directions, and that a weighted trace of the Fisher information matrix is a constant independent of
the wave function, and thus has an upper bound.
(Luo did not consider the possibility of generating prior probability distributions by normalizing the volume element of the 
Fisher information metric.)

Gnutzmann and \.Zyczkowski noted that one ``is tempted to think of the Husimi 
function as a probability density on the phase space. However, the rules for
calculating expectation values of some observable using the Husimi function
are non-classical'' \cite[sec. 2.1]{gnutzmann} (cf. \cite[p. 548]{esposito}).
Gardiner and Zoller remarked that the ``main problem of the Q-function
is that not all positive normalizable Q-functions correspond to positive
normalizable density operators'' \cite[p. 109]{crispin}.

Further, the comparison of 
 distances between Husimi distributions for arbitrary quantum states based on the Fisher information metric with those
employing the Monge distance \cite{monge}, might be investigated. 
For the TLQS studied here, the Monge distance is, in fact, ``consistent with
the geometry of the Bloch ball induced by the Hilbert-Schmidt or the trace
distance'' \cite[p. 6716]{monge}. (The trace distance is monotone, but 
{\it not}
Riemannian, while the Hilbert-Schmidt distance, contrastingly,  is Riemannian, but {\it not} monotone \cite[p. 10083]{hansjuergen} \cite{ozawa}.) For $n$-dimensional quantum
systems ($n>2$), unlike the trace, 
Hilbert-Schmidt or Bures distance, the Monge distance 
of $\rho$ to 
the fully mixed  state --- which provides information concerning 
the {\it localization} of $\rho$ in the classical phase 
space --- is {\it not} the same for all pure states
\cite{monge}.
The only monotone metrics for which explicit distance formulas are 
so-far 
available are the Bures (minimal monotone) and Wigner-Yanase ones \cite{pablo}.

\begin{figure}
\includegraphics{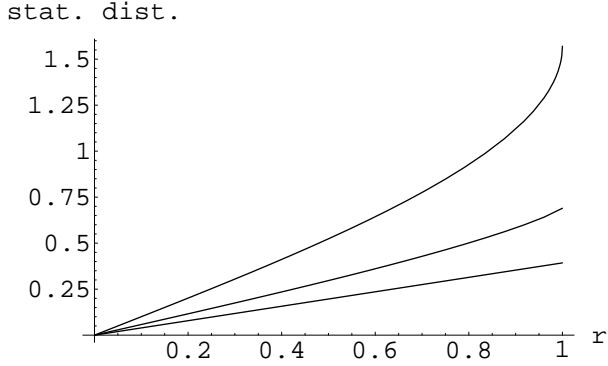}
\caption{\label{fig:SD}Statistical distance as a function of distance from the origin of the Bloch ball --- corresponding to the fully 
mixed state --- for any monotone metric, for
$ds^2_{Fisher_{Hus}}$, and for the Monge (or equivalently, for $n=2$, Hilbert-Schmidt) metric. The monotone-metric curve dominates that for $ds^2_{Fisher_{Hus}}$, which dominates the linear curve for the Monge metric.}
\end{figure}

In Fig.~\ref{fig:SD} we show how the distance from
the fully mixed state ($r=0$) increases as $r$ increases, for any
monotone metric and for $ds^2_{Fisher_{Hus}}$, and (linearly) for the Monge 
(or Hilbert-Schmidt) metric. The first-mentioned
distance --- taking the
functional form $\arcsin{r}$ (equalling $\frac{\pi}{2}$ for $r=1$) --- dominates the second-mentioned distance (equalling $\frac{\pi}{4.5551532167057}$ for 
$r=1$),
which in turns dominates the third \cite[eq. (4.10)]{monge}, which takes the value $\frac{\pi}{8}$ for $r=1$.

Let us bring to the attention of the reader, a recent preprint,
which introduces a concept of escort {\it density operators}
and a related one 
of {\it generalized} Fisher information \cite{naudts} 
(cf. \cite{lutwak,naudts2}).

We have been consistently able above to find
(apparently operator) monotone
functions to generate the tangential components of (classical) Fisher information metrics for (rotationally-symmetric) probability distributions over the
TLQS. We  suspect the existence of some (yet not formally demonstrated)
theorem to this effect.
Also, it would be of interest to formally test the various monotone functions
presented above for the property (requisite for a {\it quantum} monotone
metric \cite{petzsudar,petzLIN} of {\it operator} monotonicity.

We have ``hybridized'' $ds^2_{Fisher_{Hus}}$ above to a (quantum) monotone metric $ds^2_{HYB_{Hus}}$ by replacing its radial component by that required
($\frac{1}{1-r^2}$), while retaining its tangential component (formed from
$f_{Hus}(t)$). But it appears that we could also convert it by appropriately
scaling (a {\it conformal} transformation) 
the entire metric (tangential {\it and} radial components)
by some suitable function.
If we do so, we find that --- by explicit construction --- the new metric
($ds^2_{conformal_{Hus}}$) has the required radial component, while the tangential component is
generated by a function
\begin{equation}
f_{conformal_{Hus}}(t) = f_{Hus}(t) -t -1 ,
\end{equation}
which also appears to be operator monotone. (We note that $f_{conformal_{Hus}}(1)=1$
and $\lim_{t\to0} f_{conformal_{Hus}}(t)=0$.)
But now, we have the large relative entropies $S_{KL}(p_{GKS},p_{conformal_{Hus}}) = 
50.4636$ and $S_{KL}(p_{conformal_{Hus}},p_{GKS}) = 54.2601$.
At $r=0$, $ds^2_{conformal_{Hus}}$ is not flat, as is $ds^2_{Fisher_{Hus}}$,
but has a (limiting) scalar curvature of $-\frac{24}{5}$.
\subsection{Further questions}
Motivated by the analyses above, we would like to pose the question of whether
there exists a family of trivariate 
{\it probability} distributions parametrized by the points of the Bloch ball, for which the associated (classically 
{\it unique} [up to a constant multiple]) 
Fisher information metric
{\it fully} --- both in terms of tangential {\it and} radial 
components -- has the requisite form (\ref{radtang}) for a monotone metric.
Also, the {\it volume elements} (and hence associated 
prior probabilities) of the monotone metrics are expressible as the {\it product} of Haar measure and measures over the eigenvalues 
\cite{mjwhall}. To what extent, if any,
does this hold true for prior probabilities {\it not} arising from monotone metrics? 
Are there any non-monotone metrics which give rise to prior probabilities
{\it more} noninformative than (at the very least) the minimal monotone
(Bures) one? What are suitable counterparts to formula (\ref{radtang})
for $n$-level quantum systems ($n>2$)? Are there any monotone 
metrics which are flat at the fully mixed state, as is $ds^2_{Fisher_{Hus}}$
 (\ref{Flat})?

\section{Summary}
In a {\it classical} context,
for the family of
Husimi {\it probability} distributions
 over the three-dimensional
Bloch ball of two-level quantum systems (TLQS),  we derived the
(flat-at-the-fully-mixed-state)
 Fisher information metric ($ds^2_{Fisher_{Hus}}$, given by 
(\ref{expressionHus})). Its
tangential --- but {\it not} its radial ($r$) --- component
conformed to that of one of
the (uncountably) {\it infinite} class of (quantum) monotone metrics.
The {\it prior} probability distribution ($p_{Hus}$)
formed by normalizing the volume
element  of $ds^2_{Fisher_{Hus}}$ was found (sec.~\ref{Buresprior}) 
to be 
considerably {\it less} noninformative than
the priors formed from
 {\it any} of the (quantum) 
monotone metrics, even that ($p_{Bures}$)
based on the
(relatively informative)
{\it minimal} monotone (Bures) metric. However, if we replaced
the radial component
of $ds^2_{Fisher_{Hus}}$ by that required ($\frac{1}{1-r^2}$)
of {\it all} (quantum) 
monotone metrics, the resultant  ``hybridized-Husimi''
prior probability ($p_{HYB_{Hus}}$)
 became 
very close (in the sense of relative entropy $\approx .006$ ``nats'')
to $p_{Bures}$, and thus comparably informative in nature, but even
nearer  ($\approx .0004$)
to another quantum-monotone-metric-based
(``Grosse-Krattenthaler-Slater'' or ``quasi-Bures'') probability
distribution ($p_{GKS}$) that has been conjectured to yield the
asymptotic minimax/maximin redundancy for universal {\it quantum} coding.
The analogous (Bayesian)
role in universal (classical) coding --- by a well-known
result of Clarke and Barron \cite{clarkebarron,clarkebarron2} --- is 
played by Jeffreys'
prior (cf. \cite{slaterJeff,Kwek}). 
This takes the specific (original, {\it non}-hybridized) form $p_{Hus}$
for the family (manifold) of 
trivariate Husimi qubit probability distributions under
study. We also studied the Fisher information metric for the
{\it escort}-Husimi (sec.~\ref{EscortSec}), positive-P 
(sec.~\ref{PositiveSec}) and certain Gaussian
probability distributions (sec~\ref{GaussianSec}), 
as well as, in some sense, 
the discrete Wigner pseudoprobability (sec.~\ref{WignerSec}).
Additionally, we applied the Clarke comparative noninformativity test 
\cite{compnoninform,clarke} to quantum priors 
(sec.~\ref{seccompnon}). Evidence that this test 
is consistent with the recently-stated criterion of ``biasedness to
pure states'' of Srednicki \cite{srednicki} has been presented
\cite{slatercanosa}.

\begin{acknowledgments}
I wish to express gratitude to the Kavli Institute for Theoretical
Physics (KITP)
for computational support in this research and to C. Krattenthaler for
deriving the expressions (\ref{ck}) and (\ref{Pochhammer}) 
and to M. Trott for assistance with
certain MATHEMATICA computations.

\end{acknowledgments}

\bibliography{RevisedHus}

\begin{thebibliography}{69}
\expandafter\ifx\csname natexlab\endcsname\relax\def\natexlab#1{#1}\fi
\expandafter\ifx\csname bibnamefont\endcsname\relax
  \def\bibnamefont#1{#1}\fi
\expandafter\ifx\csname bibfnamefont\endcsname\relax
  \def\bibfnamefont#1{#1}\fi
\expandafter\ifx\csname citenamefont\endcsname\relax
  \def\citenamefont#1{#1}\fi
\expandafter\ifx\csname url\endcsname\relax
  \def\url#1{\texttt{#1}}\fi
\expandafter\ifx\csname urlprefix\endcsname\relax\def\urlprefix{URL }\fi
\providecommand{\bibinfo}[2]{#2}
\providecommand{\eprint}[2][]{\url{#2}}

\bibitem[{\citenamefont{Mosseri and Dandoloff}(2001)}]{mosseri}
\bibinfo{author}{\bibfnamefont{R.}~\bibnamefont{Mosseri}} \bibnamefont{and}
  \bibinfo{author}{\bibfnamefont{R.}~\bibnamefont{Dandoloff}},
  \bibinfo{journal}{J. Phys. A.} \textbf{\bibinfo{volume}{34}},
  \bibinfo{pages}{10243} (\bibinfo{year}{2001}).

\bibitem[{\citenamefont{Husimi}(1940)}]{khusimi}
\bibinfo{author}{\bibfnamefont{K.}~\bibnamefont{Husimi}},
  \bibinfo{journal}{Proc. Phys. Soc. Japan} \textbf{\bibinfo{volume}{22}},
  \bibinfo{pages}{264} (\bibinfo{year}{1940}).

\bibitem[{\citenamefont{\.Zyczkowski and S{\l}omczy\'nski}(2001)}]{monge}
\bibinfo{author}{\bibfnamefont{K.}~\bibnamefont{\.Zyczkowski}}
  \bibnamefont{and}
  \bibinfo{author}{\bibfnamefont{W.}~\bibnamefont{S{\l}omczy\'nski}},
  \bibinfo{journal}{J. Phys. A} \textbf{\bibinfo{volume}{34}},
  \bibinfo{pages}{6689} (\bibinfo{year}{2001}).

\bibitem[{\citenamefont{Guiasu}(1987)}]{guiasu}
\bibinfo{author}{\bibfnamefont{S.}~\bibnamefont{Guiasu}},
  \bibinfo{journal}{Phys. Rev. A} \textbf{\bibinfo{volume}{36}},
  \bibinfo{pages}{1971} (\bibinfo{year}{1987}).

\bibitem[{\citenamefont{Levine}(1988)}]{levine}
\bibinfo{author}{\bibfnamefont{R.~D.} \bibnamefont{Levine}},
  \bibinfo{journal}{J. Statist. Phys.} \textbf{\bibinfo{volume}{52}},
  \bibinfo{pages}{1203} (\bibinfo{year}{1988}).

\bibitem[{\citenamefont{Bach}(1979)}]{bach}
\bibinfo{author}{\bibfnamefont{A.}~\bibnamefont{Bach}}, \bibinfo{journal}{Phys.
  Lett. A} \textbf{\bibinfo{volume}{73}}, \bibinfo{pages}{287}
  (\bibinfo{year}{1979}).

\bibitem[{\citenamefont{Bach}(1980)}]{bach2}
\bibinfo{author}{\bibfnamefont{A.}~\bibnamefont{Bach}}, \bibinfo{journal}{J.
  Math. Phys.} \textbf{\bibinfo{volume}{21}}, \bibinfo{pages}{789}
  (\bibinfo{year}{1980}).

\bibitem[{\citenamefont{Bach}(1981)}]{bach3}
\bibinfo{author}{\bibfnamefont{A.}~\bibnamefont{Bach}}, \bibinfo{journal}{J.
  Phys. A} \textbf{\bibinfo{volume}{14}}, \bibinfo{pages}{125}
  (\bibinfo{year}{1981}).

\bibitem[{\citenamefont{Gibbons}(1992)}]{gibbons}
\bibinfo{author}{\bibfnamefont{G.~W.} \bibnamefont{Gibbons}},
  \bibinfo{journal}{J. Geom. Phys.} \textbf{\bibinfo{volume}{8}},
  \bibinfo{pages}{147} (\bibinfo{year}{1992}).

\bibitem[{\citenamefont{Sugita and Aiba}(2002)}]{sugita2}
\bibinfo{author}{\bibfnamefont{A.}~\bibnamefont{Sugita}} \bibnamefont{and}
  \bibinfo{author}{\bibfnamefont{H.}~\bibnamefont{Aiba}},
  \bibinfo{journal}{Phys. Rev. E} \textbf{\bibinfo{volume}{65}},
  \bibinfo{pages}{036205} (\bibinfo{year}{2002}).

\bibitem[{\citenamefont{Penrose}(2005)}]{penrose}
\bibinfo{author}{\bibfnamefont{R.}~\bibnamefont{Penrose}},
  \emph{\bibinfo{title}{The Road to Reality}} (\bibinfo{publisher}{A. A.
  Knopf}, \bibinfo{address}{New York}, \bibinfo{year}{2005}).

\bibitem[{\citenamefont{Chentsov}(1982)}]{chentsov}
\bibinfo{author}{\bibfnamefont{N.~N.} \bibnamefont{Chentsov}},
  \emph{\bibinfo{title}{Statistical Decision Rules and Optimal Inference}}
  (\bibinfo{publisher}{Amer. Mat. Soc.}, \bibinfo{address}{Providence},
  \bibinfo{year}{1982}).

\bibitem[{\citenamefont{Campbell}(1986)}]{campbell}
\bibinfo{author}{\bibfnamefont{L.~L.} \bibnamefont{Campbell}},
  \bibinfo{journal}{Proc. Amer. Math. Soc.} \textbf{\bibinfo{volume}{98}},
  \bibinfo{pages}{135} (\bibinfo{year}{1986}).

\bibitem[{\citenamefont{Frieden}(2004)}]{frieden}
\bibinfo{author}{\bibfnamefont{B.~R.} \bibnamefont{Frieden}},
  \emph{\bibinfo{title}{Science from Fisher Information: A Unification}}
  (\bibinfo{publisher}{Cambridge Univ}, \bibinfo{address}{Cambridge},
  \bibinfo{year}{2004}).

\bibitem[{\citenamefont{Petz and Sud\mbox{\'a}r}(1996)}]{petzsudar}
\bibinfo{author}{\bibfnamefont{D.}~\bibnamefont{Petz}} \bibnamefont{and}
  \bibinfo{author}{\bibfnamefont{C.}~\bibnamefont{Sud\mbox{\'a}r}},
  \bibinfo{journal}{J. Math. Phys.} \textbf{\bibinfo{volume}{37}},
  \bibinfo{pages}{2662} (\bibinfo{year}{1996}).

\bibitem[{\citenamefont{Hall}(1998)}]{mjwhall}
\bibinfo{author}{\bibfnamefont{M.~J.~W.} \bibnamefont{Hall}},
  \bibinfo{journal}{Phys.Lett.A} \textbf{\bibinfo{volume}{242}},
  \bibinfo{pages}{123} (\bibinfo{year}{1998}).

\bibitem[{\citenamefont{Clarke and Barron}(1990)}]{clarkebarron}
\bibinfo{author}{\bibfnamefont{B.~S.} \bibnamefont{Clarke}} \bibnamefont{and}
  \bibinfo{author}{\bibfnamefont{A.~R.} \bibnamefont{Barron}},
  \bibinfo{journal}{IEEE Trans. Info. Theory} \textbf{\bibinfo{volume}{36}},
  \bibinfo{pages}{453} (\bibinfo{year}{1990}).

\bibitem[{\citenamefont{Clarke and Barron}(1994)}]{clarkebarron2}
\bibinfo{author}{\bibfnamefont{B.~S.} \bibnamefont{Clarke}} \bibnamefont{and}
  \bibinfo{author}{\bibfnamefont{A.~R.} \bibnamefont{Barron}},
  \bibinfo{journal}{J. Statist. Plann. Inference}
  \textbf{\bibinfo{volume}{41}}, \bibinfo{pages}{37} (\bibinfo{year}{1994}).

\bibitem[{\citenamefont{Slater}(1996{\natexlab{a}})}]{slaterJeff}
\bibinfo{author}{\bibfnamefont{P.~B.} \bibnamefont{Slater}},
  \bibinfo{journal}{J. Phys. A} \textbf{\bibinfo{volume}{29}},
  \bibinfo{pages}{L601} (\bibinfo{year}{1996}{\natexlab{a}}).

\bibitem[{\citenamefont{Kwek et~al.}(1999)\citenamefont{Kwek, Oh, and
  Wang}}]{Kwek}
\bibinfo{author}{\bibfnamefont{L.~C.} \bibnamefont{Kwek}},
  \bibinfo{author}{\bibfnamefont{C.~H.} \bibnamefont{Oh}}, \bibnamefont{and}
  \bibinfo{author}{\bibfnamefont{X.-B.} \bibnamefont{Wang}},
  \bibinfo{journal}{J. Phys. A} \textbf{\bibinfo{volume}{32}},
  \bibinfo{pages}{6613} (\bibinfo{year}{1999}).

\bibitem[{\citenamefont{Petz}(1996)}]{petzLIN}
\bibinfo{author}{\bibfnamefont{D.}~\bibnamefont{Petz}},
  \bibinfo{journal}{Linear Alg. Applics.} \textbf{\bibinfo{volume}{244}},
  \bibinfo{pages}{81} (\bibinfo{year}{1996}).

\bibitem[{\citenamefont{Slater}(2000)}]{slaterlmp}
\bibinfo{author}{\bibfnamefont{P.~B.} \bibnamefont{Slater}},
  \bibinfo{journal}{Lett. Math. Phys.} \textbf{\bibinfo{volume}{52}},
  \bibinfo{pages}{343} (\bibinfo{year}{2000}).

\bibitem[{\citenamefont{Petz}(2002)}]{petzJPA}
\bibinfo{author}{\bibfnamefont{D.}~\bibnamefont{Petz}}, \bibinfo{journal}{J.
  Phys. A} \textbf{\bibinfo{volume}{35}}, \bibinfo{pages}{929}
  (\bibinfo{year}{2002}).

\bibitem[{\citenamefont{Luo}(2001)}]{luo}
\bibinfo{author}{\bibfnamefont{S.}~\bibnamefont{Luo}}, \bibinfo{journal}{J.
  Statist. Phys.} \textbf{\bibinfo{volume}{102}}, \bibinfo{pages}{1417}
  (\bibinfo{year}{2001}).

\bibitem[{\citenamefont{Slater}(1998)}]{compnoninform}
\bibinfo{author}{\bibfnamefont{P.~B.} \bibnamefont{Slater}},
  \bibinfo{journal}{Phys. Lett. A} \textbf{\bibinfo{volume}{247}},
  \bibinfo{pages}{1} (\bibinfo{year}{1998}).

\bibitem[{\citenamefont{Clarke}(1996)}]{clarke}
\bibinfo{author}{\bibfnamefont{B.}~\bibnamefont{Clarke}}, \bibinfo{journal}{J.
  Amer. Statist. Assoc.} \textbf{\bibinfo{volume}{91}}, \bibinfo{pages}{173}
  (\bibinfo{year}{1996}).

\bibitem[{\citenamefont{Srednicki}(2005)}]{srednicki}
\bibinfo{author}{\bibfnamefont{M.}~\bibnamefont{Srednicki}},
  \bibinfo{journal}{Phys. Rev. A} \textbf{\bibinfo{volume}{71}},
  \bibinfo{pages}{052107} (\bibinfo{year}{2005}).

\bibitem[{\citenamefont{Slater}()}]{slatercanosa}
\bibinfo{author}{\bibfnamefont{P.~B.} \bibnamefont{Slater}},
  \eprint{quant-ph/0507203 (to appear in J. Phys. A)}.

\bibitem[{\citenamefont{Slater}(1996{\natexlab{b}})}]{slaterSPIN}
\bibinfo{author}{\bibfnamefont{P.~B.} \bibnamefont{Slater}},
  \bibinfo{journal}{J. Phys. A} \textbf{\bibinfo{volume}{29}},
  \bibinfo{pages}{L271} (\bibinfo{year}{1996}{\natexlab{b}}).

\bibitem[{\citenamefont{Slater}(1996{\natexlab{c}})}]{slaterSPIN2}
\bibinfo{author}{\bibfnamefont{P.~B.} \bibnamefont{Slater}},
  \bibinfo{journal}{J.Math. Phys.} \textbf{\bibinfo{volume}{37}},
  \bibinfo{pages}{2682} (\bibinfo{year}{1996}{\natexlab{c}}).

\bibitem[{\citenamefont{Mintert and \.Zyczkowski}(2004)}]{mintert}
\bibinfo{author}{\bibfnamefont{F.}~\bibnamefont{Mintert}} \bibnamefont{and}
  \bibinfo{author}{\bibfnamefont{K.}~\bibnamefont{\.Zyczkowski}},
  \bibinfo{journal}{Phys. Rev. A} \textbf{\bibinfo{volume}{69}},
  \bibinfo{pages}{022317} (\bibinfo{year}{2004}).

\bibitem[{\citenamefont{Lesniewski and Ruskai}(1999)}]{lesniewski}
\bibinfo{author}{\bibfnamefont{A.}~\bibnamefont{Lesniewski}} \bibnamefont{and}
  \bibinfo{author}{\bibfnamefont{M.~B.} \bibnamefont{Ruskai}},
  \bibinfo{journal}{J. Math. Phys.} \textbf{\bibinfo{volume}{40}},
  \bibinfo{pages}{5702} (\bibinfo{year}{1999}).

\bibitem[{\citenamefont{Jen\u{c}ov\'a}(2004)}]{jencova}
\bibinfo{author}{\bibfnamefont{A.}~\bibnamefont{Jen\u{c}ov\'a}},
  \bibinfo{journal}{Intl. J. Theor. Phys.} \textbf{\bibinfo{volume}{43}},
  \bibinfo{pages}{1635} (\bibinfo{year}{2004}).

\bibitem[{\citenamefont{Luo}(2003)}]{luoWY}
\bibinfo{author}{\bibfnamefont{S.}~\bibnamefont{Luo}}, \bibinfo{journal}{Proc.
  Amer. Math. Soc.} \textbf{\bibinfo{volume}{132}}, \bibinfo{pages}{885}
  (\bibinfo{year}{2003}).

\bibitem[{\citenamefont{Slater}(2001)}]{slaterefficiency}
\bibinfo{author}{\bibfnamefont{P.~B.} \bibnamefont{Slater}},
  \bibinfo{journal}{J. Phys. A} \textbf{\bibinfo{volume}{34}},
  \bibinfo{pages}{7029} (\bibinfo{year}{2001}).

\bibitem[{\citenamefont{Decker et~al.}(2004)\citenamefont{Decker, Janzing, and
  Beth}}]{decker}
\bibinfo{author}{\bibfnamefont{T.}~\bibnamefont{Decker}},
  \bibinfo{author}{\bibfnamefont{D.}~\bibnamefont{Janzing}}, \bibnamefont{and}
  \bibinfo{author}{\bibfnamefont{T.}~\bibnamefont{Beth}},
  \bibinfo{journal}{Intl. J. Quant. Inform.} \textbf{\bibinfo{volume}{2}},
  \bibinfo{pages}{353} (\bibinfo{year}{2004}).

\bibitem[{\citenamefont{Ozawa}(2001)}]{ozawa}
\bibinfo{author}{\bibfnamefont{M.}~\bibnamefont{Ozawa}},
  \bibinfo{journal}{Phys. Lett. A} \textbf{\bibinfo{volume}{268}},
  \bibinfo{pages}{158} (\bibinfo{year}{2001}).

\bibitem[{\citenamefont{Krattenthaler and Slater}(2000)}]{krattenthaler}
\bibinfo{author}{\bibfnamefont{C.}~\bibnamefont{Krattenthaler}}
  \bibnamefont{and} \bibinfo{author}{\bibfnamefont{P.~B.}
  \bibnamefont{Slater}}, \bibinfo{journal}{IEEE Trans. Info. Theory}
  \textbf{\bibinfo{volume}{46}}, \bibinfo{pages}{801} (\bibinfo{year}{2000}).

\bibitem[{\citenamefont{Hayashi and Matsumoto}(2002)}]{hayashi}
\bibinfo{author}{\bibfnamefont{M.}~\bibnamefont{Hayashi}} \bibnamefont{and}
  \bibinfo{author}{\bibfnamefont{K.}~\bibnamefont{Matsumoto}},
  \bibinfo{journal}{Phys. Rev. A} \textbf{\bibinfo{volume}{66}},
  \bibinfo{pages}{022311} (\bibinfo{year}{2002}).

\bibitem[{\citenamefont{Jozsa and Presnell}(2003)}]{jozsa}
\bibinfo{author}{\bibfnamefont{R.}~\bibnamefont{Jozsa}} \bibnamefont{and}
  \bibinfo{author}{\bibfnamefont{S.}~\bibnamefont{Presnell}},
  \bibinfo{journal}{Proc. Royal Soc. Lond. A} \textbf{\bibinfo{volume}{459}},
  \bibinfo{pages}{3061} (\bibinfo{year}{2003}).

\bibitem[{\citenamefont{Jozsa et~al.}(1998)\citenamefont{Jozsa, Horodecki,
  Horodecki, and Horodecki}}]{jozsa1}
\bibinfo{author}{\bibfnamefont{R.}~\bibnamefont{Jozsa}},
  \bibinfo{author}{\bibfnamefont{M.}~\bibnamefont{Horodecki}},
  \bibinfo{author}{\bibfnamefont{P.}~\bibnamefont{Horodecki}},
  \bibnamefont{and}
  \bibinfo{author}{\bibfnamefont{R.}~\bibnamefont{Horodecki}},
  \bibinfo{journal}{Phys. Rev. Lett.} \textbf{\bibinfo{volume}{81}},
  \bibinfo{pages}{1714} (\bibinfo{year}{1998}).

\bibitem[{\citenamefont{Grosse et~al.}()\citenamefont{Grosse, Krattenthaler,
  and Slater}}]{asyred2}
\bibinfo{author}{\bibfnamefont{H.}~\bibnamefont{Grosse}},
  \bibinfo{author}{\bibfnamefont{C.}~\bibnamefont{Krattenthaler}},
  \bibnamefont{and} \bibinfo{author}{\bibfnamefont{P.~B.}
  \bibnamefont{Slater}}, \emph{\bibinfo{title}{Asymptotic redundancies for
  universal quantum coding. part 2 (in preparation)}}.

\bibitem[{\citenamefont{Caves et~al.}(2002)\citenamefont{Caves, Fuchs, and
  Schack}}]{caves}
\bibinfo{author}{\bibfnamefont{C.~M.} \bibnamefont{Caves}},
  \bibinfo{author}{\bibfnamefont{C.~A.} \bibnamefont{Fuchs}}, \bibnamefont{and}
  \bibinfo{author}{\bibfnamefont{R.}~\bibnamefont{Schack}},
  \bibinfo{journal}{J. Math. Phys.} \textbf{\bibinfo{volume}{43}},
  \bibinfo{pages}{4537} (\bibinfo{year}{2002}).

\bibitem[{\citenamefont{Pennini and Plastino}(2004)}]{pennini}
\bibinfo{author}{\bibfnamefont{F.}~\bibnamefont{Pennini}} \bibnamefont{and}
  \bibinfo{author}{\bibfnamefont{A.}~\bibnamefont{Plastino}},
  \bibinfo{journal}{Phys. Lett. A} \textbf{\bibinfo{volume}{326}},
  \bibinfo{pages}{20} (\bibinfo{year}{2004}).

\bibitem[{\citenamefont{Amari}(1982)}]{amari}
\bibinfo{author}{\bibfnamefont{S.~I.} \bibnamefont{Amari}},
  \bibinfo{journal}{Ann. Statist.} \textbf{\bibinfo{volume}{10}},
  \bibinfo{pages}{357} (\bibinfo{year}{1982}).

\bibitem[{\citenamefont{Naudts}(2004)}]{naudts2}
\bibinfo{author}{\bibfnamefont{J.}~\bibnamefont{Naudts}}, \bibinfo{journal}{J.
  Inequal. Pure Appl. Math.} \textbf{\bibinfo{volume}{5}}, \bibinfo{pages}{1}
  (\bibinfo{year}{2004}).

\bibitem[{\citenamefont{Gnutzmann and \.Zyczkowski}(2001)}]{gnutzmann}
\bibinfo{author}{\bibfnamefont{S.}~\bibnamefont{Gnutzmann}} \bibnamefont{and}
  \bibinfo{author}{\bibfnamefont{K.}~\bibnamefont{\.Zyczkowski}},
  \bibinfo{journal}{J. Phys. A} \textbf{\bibinfo{volume}{34}},
  \bibinfo{pages}{10123} (\bibinfo{year}{2001}).

\bibitem[{\citenamefont{Sugita}(2003)}]{sugita}
\bibinfo{author}{\bibfnamefont{A.}~\bibnamefont{Sugita}}, \bibinfo{journal}{J.
  Phys. A} \textbf{\bibinfo{volume}{36}}, \bibinfo{pages}{9081}
  (\bibinfo{year}{2003}).

\bibitem[{\citenamefont{Beck}(2001)}]{beck}
\bibinfo{author}{\bibfnamefont{C.}~\bibnamefont{Beck}}, \bibinfo{journal}{Phys.
  Lett. A} \textbf{\bibinfo{volume}{287}}, \bibinfo{pages}{240}
  (\bibinfo{year}{2001}).

\bibitem[{\citenamefont{Braunstein et~al.}(1991)\citenamefont{Braunstein,
  Caves, and Milburn}}]{bmc}
\bibinfo{author}{\bibfnamefont{S.~L.} \bibnamefont{Braunstein}},
  \bibinfo{author}{\bibfnamefont{C.~M.} \bibnamefont{Caves}}, \bibnamefont{and}
  \bibinfo{author}{\bibfnamefont{G.~J.} \bibnamefont{Milburn}},
  \bibinfo{journal}{Phys. Rev. A} \textbf{\bibinfo{volume}{43}},
  \bibinfo{pages}{1153} (\bibinfo{year}{1991}).

\bibitem[{\citenamefont{Gardiner and Zoller}(2004)}]{crispin}
\bibinfo{author}{\bibfnamefont{C.~W.} \bibnamefont{Gardiner}} \bibnamefont{and}
  \bibinfo{author}{\bibfnamefont{P.}~\bibnamefont{Zoller}},
  \emph{\bibinfo{title}{Quantum Noise}} (\bibinfo{publisher}{Springer},
  \bibinfo{address}{Berlin}, \bibinfo{year}{2004}).

\bibitem[{\citenamefont{Lee}(1995)}]{lee}
\bibinfo{author}{\bibfnamefont{H.~W.} \bibnamefont{Lee}},
  \bibinfo{journal}{Phys. Rep.} \textbf{\bibinfo{volume}{259}},
  \bibinfo{pages}{147} (\bibinfo{year}{1995}).

\bibitem[{\citenamefont{Slater}(1995)}]{quantumcointossing}
\bibinfo{author}{\bibfnamefont{P.~B.} \bibnamefont{Slater}},
  \bibinfo{journal}{Phys. Lett. A} \textbf{\bibinfo{volume}{206}},
  \bibinfo{pages}{66} (\bibinfo{year}{1995}).

\bibitem[{\citenamefont{Slater}(1996{\natexlab{d}})}]{physicaA}
\bibinfo{author}{\bibfnamefont{P.~B.} \bibnamefont{Slater}},
  \bibinfo{journal}{Phys. A} \textbf{\bibinfo{volume}{223}},
  \bibinfo{pages}{167} (\bibinfo{year}{1996}{\natexlab{d}}).

\bibitem[{\citenamefont{Yuen and Lax}(1973)}]{YuenLax}
\bibinfo{author}{\bibfnamefont{H.~P.} \bibnamefont{Yuen}} \bibnamefont{and}
  \bibinfo{author}{\bibfnamefont{M.}~\bibnamefont{Lax}}, \bibinfo{journal}{IEEE
  Trans. Info. Theory} \textbf{\bibinfo{volume}{19}}, \bibinfo{pages}{740}
  (\bibinfo{year}{1973}).

\bibitem[{\citenamefont{Galv{\~a}o}(2005)}]{galvao}
\bibinfo{author}{\bibfnamefont{E.~F.} \bibnamefont{Galv{\~a}o}},
  \bibinfo{journal}{Phys. Rev. A} \textbf{\bibinfo{volume}{71}},
  \bibinfo{pages}{042302} (\bibinfo{year}{2005}).

\bibitem[{\citenamefont{Asplund and Bj{\"o}rk}(2001)}]{asplund}
\bibinfo{author}{\bibfnamefont{R.}~\bibnamefont{Asplund}} \bibnamefont{and}
  \bibinfo{author}{\bibfnamefont{G.}~\bibnamefont{Bj{\"o}rk}},
  \bibinfo{journal}{Phys. Rev. A} \textbf{\bibinfo{volume}{64}},
  \bibinfo{pages}{012106} (\bibinfo{year}{2001}).

\bibitem[{\citenamefont{Cunha et~al.}(2001)\citenamefont{Cunha, Man'ko, and
  Scully}}]{cunha}
\bibinfo{author}{\bibfnamefont{M.~O.~T.} \bibnamefont{Cunha}},
  \bibinfo{author}{\bibfnamefont{V.~I.} \bibnamefont{Man'ko}},
  \bibnamefont{and} \bibinfo{author}{\bibfnamefont{M.~O.}
  \bibnamefont{Scully}}, \bibinfo{journal}{Found. Phys. Lett.} p.
  \bibinfo{pages}{103} (\bibinfo{year}{2001}).

\bibitem[{\citenamefont{Dittmann}(1999)}]{dittmannjgp}
\bibinfo{author}{\bibfnamefont{J.}~\bibnamefont{Dittmann}},
  \bibinfo{journal}{J. Geom. Phys.} \textbf{\bibinfo{volume}{31}},
  \bibinfo{pages}{16} (\bibinfo{year}{1999}).

\bibitem[{\citenamefont{Andai}(2003)}]{attila}
\bibinfo{author}{\bibfnamefont{A.}~\bibnamefont{Andai}}, \bibinfo{journal}{J.
  Math. Phys.} \textbf{\bibinfo{volume}{44}}, \bibinfo{pages}{3675}
  (\bibinfo{year}{2003}).

\bibitem[{\citenamefont{Gibilisco and Isola}(2003)}]{pablo}
\bibinfo{author}{\bibfnamefont{P.}~\bibnamefont{Gibilisco}} \bibnamefont{and}
  \bibinfo{author}{\bibfnamefont{T.}~\bibnamefont{Isola}}, \bibinfo{journal}{J.
  Math. Phys.} \textbf{\bibinfo{volume}{44}}, \bibinfo{pages}{3752}
  (\bibinfo{year}{2003}).

\bibitem[{\citenamefont{H{\"u}bner}(1992)}]{hubner}
\bibinfo{author}{\bibfnamefont{M.}~\bibnamefont{H{\"u}bner}},
  \bibinfo{journal}{Phys. Lett. A} \textbf{\bibinfo{volume}{163}},
  \bibinfo{pages}{239} (\bibinfo{year}{1992}).

\bibitem[{\citenamefont{Moln{\'a}r and Timmermann}(2003)}]{lajos}
\bibinfo{author}{\bibfnamefont{L.}~\bibnamefont{Moln{\'a}r}} \bibnamefont{and}
  \bibinfo{author}{\bibfnamefont{W.}~\bibnamefont{Timmermann}},
  \bibinfo{journal}{J. Phys. A} \textbf{\bibinfo{volume}{36}},
  \bibinfo{pages}{267} (\bibinfo{year}{2003}).

\bibitem[{\citenamefont{Gursky and LeBrun}(1998)}]{gursky}
\bibinfo{author}{\bibfnamefont{M.~J.} \bibnamefont{Gursky}} \bibnamefont{and}
  \bibinfo{author}{\bibfnamefont{C.}~\bibnamefont{LeBrun}},
  \bibinfo{journal}{GAFA Geom. Funct. Anal.} \textbf{\bibinfo{volume}{8}},
  \bibinfo{pages}{965} (\bibinfo{year}{1998}).

\bibitem[{\citenamefont{Twamley}(1996)}]{twamley}
\bibinfo{author}{\bibfnamefont{J.}~\bibnamefont{Twamley}}, \bibinfo{journal}{J.
  Phys. A} \textbf{\bibinfo{volume}{29}}, \bibinfo{pages}{3723}
  (\bibinfo{year}{1996}).

\bibitem[{\citenamefont{Esposito et~al.}(2004)\citenamefont{Esposito, Marmo,
  and Sudarshan}}]{esposito}
\bibinfo{author}{\bibfnamefont{G.}~\bibnamefont{Esposito}},
  \bibinfo{author}{\bibfnamefont{G.}~\bibnamefont{Marmo}}, \bibnamefont{and}
  \bibinfo{author}{\bibfnamefont{G.}~\bibnamefont{Sudarshan}},
  \emph{\bibinfo{title}{From Classical to Quantum Mechanics}}
  (\bibinfo{publisher}{Cambridge Univ. Press}, \bibinfo{address}{Cambridge},
  \bibinfo{year}{2004}).

\bibitem[{\citenamefont{Sommers and \.Zyczkowski}(2003)}]{hansjuergen}
\bibinfo{author}{\bibfnamefont{H.-J.} \bibnamefont{Sommers}} \bibnamefont{and}
  \bibinfo{author}{\bibfnamefont{K.}~\bibnamefont{\.Zyczkowski}},
  \bibinfo{journal}{J. Phys. A} \textbf{\bibinfo{volume}{36}},
  \bibinfo{pages}{10083} (\bibinfo{year}{2003}).

\bibitem[{\citenamefont{Naudts}(2005)}]{naudts}
\bibinfo{author}{\bibfnamefont{J.}~\bibnamefont{Naudts}},
  \bibinfo{journal}{Open Syst. Inform. Dyn.} \textbf{\bibinfo{volume}{12}},
  \bibinfo{pages}{13} (\bibinfo{year}{2005}).

\bibitem[{\citenamefont{Lutwak et~al.}(2005)\citenamefont{Lutwak, Yang, and
  Zhang}}]{lutwak}
\bibinfo{author}{\bibfnamefont{E.}~\bibnamefont{Lutwak}},
  \bibinfo{author}{\bibfnamefont{D.}~\bibnamefont{Yang}}, \bibnamefont{and}
  \bibinfo{author}{\bibfnamefont{G.~Y.} \bibnamefont{Zhang}},
  \bibinfo{journal}{IEEE Trans. Info. Theory} \textbf{\bibinfo{volume}{51}},
  \bibinfo{pages}{473} (\bibinfo{year}{2005}).

\end{thebibliography}

\end{document}